\documentclass[twocolumn,pra,amssymb,showpacs]{revtex4-1}
\usepackage{graphicx}
\usepackage{amsmath}
\usepackage{subfigure}

\newcommand{\be}{\begin{equation}}
\newcommand{\ee}{\end{equation}}
\newcommand{\ben}{\begin{eqnarray}}
\newcommand{\een}{\end{eqnarray}}

\begin{document}

\title{Computing the maximum violation of a Bell inequality is NP-complete}

\author{J. Batle$^{1}$ and C. H. Raymond Ooi$^{2}$}
\email{E-mail address: jbv276@uib.es}
\affiliation{ 
$^1$Departament de F\'{\i}sica, Universitat de les Illes Balears, 07122 Palma de Mallorca, Balearic Islands, Europe \\
$^2$Department of Physics, University of Malaya, 50603 Kuala Lumpur, Malaysia \\\\}

\date{\today}

\begin{abstract}

The number of steps required in order to maximize a Bell inequality for arbitrary number of qubits is shown to grow exponentially with either the number of steps and the number of parties involved. The 
proof that the optimization of such correlation measure is a NP-problem is based on an operational perspective involving a Turing machine, which follows a general algorithm. The implications for the computability of the so called {\it nonlocality} for any number of qubits is similar to recent results involving entanglement or similar quantum correlation-based measures.  

\end{abstract}


\maketitle

\section{Introduction}

Quantum correlations lie at the heart of quantum information theory. They are responsible for some tasks that posses no classical counterpart. It is plain from the fact that quantum measures are 
essential feature for quantum computation or secure quantum
communication, that one has to be able to develop some procedures (physical
or purely mathematical in origin) so as to ascertain whether the state $\rho$
representing the physical system under consideration is appropriate for
developing a given non-classical task. Among those correlations, 
entanglement is perhaps one of the most fundamental and non-classical features exhibited by quantum
systems \cite{LPS98}, that lies at the basis of some of the most important
processes studied by quantum information theory \cite{Galindo,NC00,LPS98,WC97,W98}
such as quantum cryptographic key distribution \cite{E91}, quantum
teleportation \cite{BBCJPW93}, superdense coding \cite{BW93}, and quantum
computation \cite{EJ96,BDMT98}. 

Other measures have been introduced in the literature that grasp features that are not captured by entanglement. They are not directly related to 
entanglement, but in some cases --specially when dealing with systems of qubits greater 
that two-- they provide a satisfactory approximate answer, like the maximum violation 
of a Bell inequality, that is, nonlocality. Local Variable Models (LVM) cannot exhibit arbitrary correlations. 
Mathematically, the conditions these correlations
must obey can always be written as inequalities --the Bell inequalities-- satisfied for the joint
probabilities of outcomes. We say that a quantum state $\rho$ is nonlocal if and only if there
are measurements on $\rho$ that produce a correlation that violates a Bell inequality. 

Later work by Zurek and Ollivier \cite{olli} established that not even
entanglement captures all aspects of quantum correlations. These
authors introduced an information-theoretical measure, quantum
discord, that corresponds to a new facet of the ``quantumness"
that arises even for non-entangled states. Indeed, it turned out
that the vast majority of quantum states exhibit a finite amount
of quantum discord. Besides its intrinsic
conceptual interest, the study of quantum discord may also have
technological implications: examples of improved quantum computing
tasks that take advantage of quantum correlations but do not rely
on entanglement have been reported [see for instance, among a
quite extensive references-list \cite{geom,olli,ferraro,dattaprl,luo}]. 
Actually, in some cases entangled states are useful to solve a problem if and only if they
violate a Bell inequality \cite{comcomplex}. Moreover, there are
important instances of non-classical information tasks that are
based directly upon non-locality, with no explicit reference to
the quantum mechanical formalism or to the associated concept of
entanglement \cite{device}. A recent work studying how entanglement can be estimated from a Bell inequality violation also sheds new light on the use of Bell inequalities \cite{EntanglementFromBell}

In any case, the study of entanglement in multipartite quantum systems has been limited to few cases. As a consequence, other measures have been introduced in order the describe the ``quantumness'' of a certain (usually mixed) state $\rho$. In recent years the use of the maximum violation of a Bell inequality serves the purpose of describing how {\it nonlocal} the state of the system is (see {\cite{Mauro}} and references therein). Although there is little connection between entanglement and nonlocality (the former is based on how the tensor structure of the concomitant Hilbert space is split, whereas the latter ascertains how well a LVM can mimic quantum mechanics), nonlocality is a good candidate for describing correlations in quantum systems. 

To be more precise, {\it the maximum violation of a Bell inequality for $N$ parties} $B_{N}^{\max}$ is the quantity chosen to approach entanglement is those scenarios \cite{jo1,jo2}. Thus, to know whether the
computation of the maximum value of a Bell inequality is NP-hard seems a relevant and reasonable question. Previous approaches in the literature have dealt with the simplest possible instance of two parties \cite{Pitowsky}, Alice and Bob, each possessing two nearly dichotomic observables. In the case of the Clauser-Horne-Shimony-Holt Bell inequality (CHSH) \cite{CHSH} , which is the strongest possible inequality for two parties (two qubits), it was proved that its maximum violation requires a computational work which grows exponentially with the number of steps required, that is, it is a NP-problem \cite{Pitowsky}.

In addition, a quantum measure such as discord has been recently proved to be NP-complete for the case of two qubits \cite{Berkeley}. The fact that the computation of some entanglement and correlated quantities is NP-hard is usually a consequence of the optimization involved in the definitions. 
The traditional tools required for optimizing Bell inequalities are borrowed from linear programming: inequalities are translated into convex polytopes, usually in high dimensions, and the proof for the general case of $N$ parties involves a gigantic task which has not been successfully solved to date.

The purpose of the present work is to provide an operational approach to the process of carrying out the maximization of a Bell inequality, based on a Turing machine, which will prove to be a NP-problem. A key ingredient will be the fact that Bell inequalities (not for probabilities) possess a recursive expression when they are generalized to $N$ qubits. In Section II we review previous results for CHSH for two qubits. In Section III we introduce the structure of the Bell inequalities employed, as well as the algorithm that the Turing machine will perform in this scenario. Finally, some conclusions are drawn in Section IV.

\section{Previous results}

The first approach to a Bell inequality (the CHSH in this case) for two qubits was carried out by Pitowsky \cite{Pitowsky}. He carries out several extremely interesting investigations concerning the foundations of quantum mechanics. He also brings together high level characterizations,
in geometrical language, of allowed classical and quantum correlation patterns. 

By using ``classical correlation polytopes'', he provides significant insight
into familiar the CHSH Bell-type inequalities. Pitowsky provides an algorithm for finding the set of ``generalized Bell inequalities'' corresponding to any particular choice of the vectors defining the hyperplanes of a convex polytope. Afterwards he proves it to be a not efficient one, that is, NP-complete. Thus, already for $N=2$ qubits, the procedure of obtaining the maximum violation of the CHSH Bell inequality is inefficient.

\begin{figure}[htbp]
\begin{center}
\includegraphics[width=8.8cm]{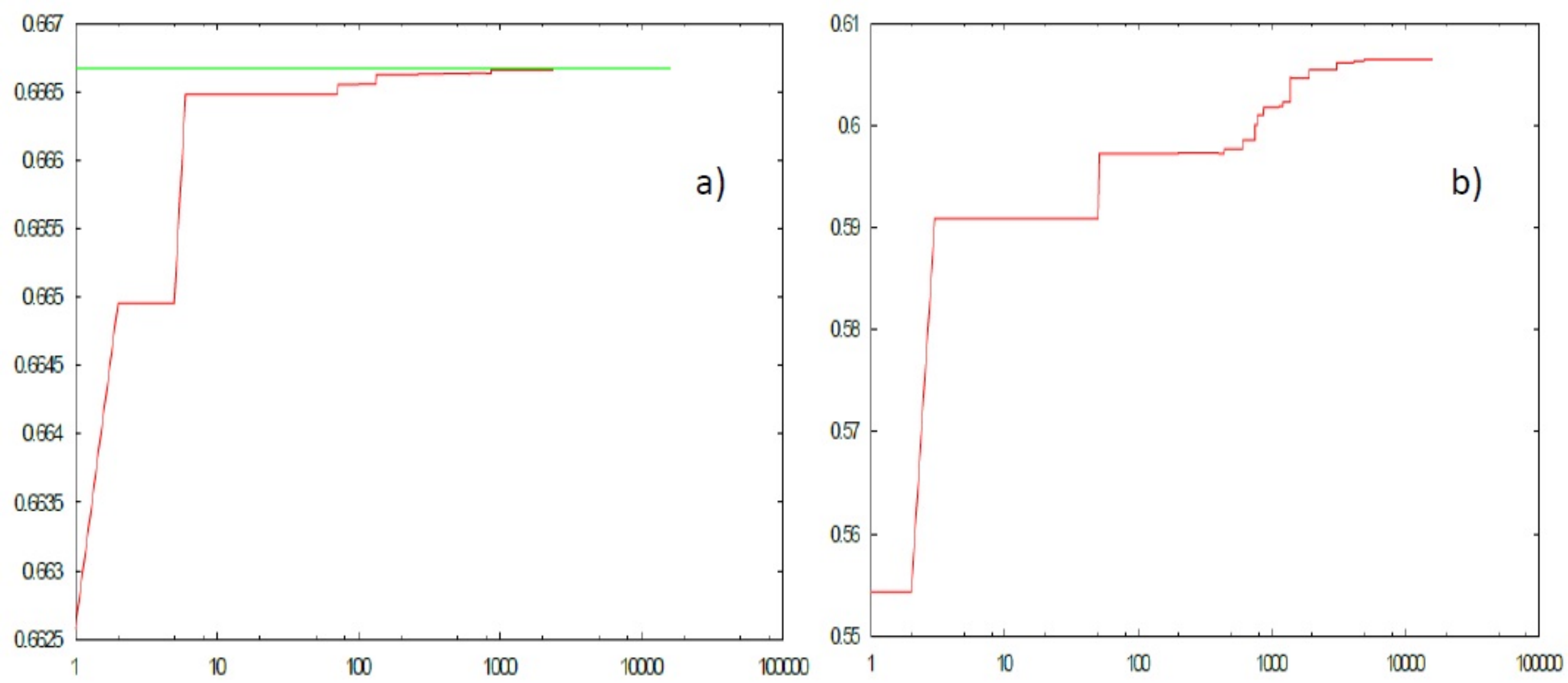}
\caption{(Color online) Plot of the typical performance of a simulated annealing maximizing the CHSH 
Bell inequality for a random state $\rho$ of two qubits vs the number of Monte Carlo steps (MC). Every MC step contains 1000 random runs. Fig. (a) depicts the evolution of the maximum of CHSH for the state 
$\frac{1}{3}|00 \rangle \langle 00|+\frac{1}{3}|01 \rangle \langle 01|+\frac{1}{3}|11 \rangle \langle 11|$, which is analytic (2/3). Fig. (b) depicts the same quantity for a random state $\rho$. 
See text for details.}
\label{fig_line}
\end{center}
\end{figure}

\section{The Turing machine and the generalized optimization of the Bell inequality}

\subsection{Bell inequalities}

Most of our knowledge on Bell inequalities and their quantum mechanical violation is based
on the CHSH inequality \cite{CHSH}. With two dichotomic observables per party, it is the
simplest \cite{Collins} (up to local symmetries) nontrivial Bell inequality for the bipartite case with
binary inputs and outcomes. Let $A_1$ and $A_2$ be two possible measurements on A side whose
outcomes are $a_j\in \lbrace -1,+1\rbrace$, and similarly for the B side. Mathematically, it can
be shown that, following LVM,
$|{\cal B}_{CHSH}^{LVM}(\lambda)|=|a_1b_1+a_1b_2+a_2b_1-a_2b_2|\leq 2$. Since $a_1$($b_1$)
and $a_2$($b_2$) cannot be measured simultaneously, instead one estimates after randomly
chosen measurements the average value ${\cal B}_{CHSH}^{LVM} \equiv \sum_{\lambda} {\cal B}_{CHSH}^{LVM}(\lambda) \mu(\lambda)=
E(A_1,B_1)+E(A_1,B_2)+E(A_2,B_1)-E(A_2,B_2)$, where $E(\cdot)$ represents the expectation value.
Therefore the CHSH inequality reduces to

\begin{equation} \label{CHSH_LVM}
|{\cal B}_{CHSH}^{LVM}| \leq 2.
\end{equation}

Quantum mechanically, since we are dealing with qubits, these observables reduce
to ${\bf A_j}({\bf B_j})={\bf a_j}({\bf b_j}) \cdot {\bf \sigma}$, where ${\bf a_j}({\bf b_j})$
are unit vectors in $\mathbb{R}^3$ and ${\bf \sigma}=(\sigma_x,\sigma_y,\sigma_z)$ are the usual
Pauli matrices. Therefore the quantal prediction for (\ref{CHSH_LVM}) reduces to the expectation
value of the operator ${\cal B}_{CHSH}$

\begin{equation} \label{CHSH_QM}
{\bf A_1}\otimes {\bf B_1} + {\bf A_1}\otimes {\bf B_2}
+ {\bf A_2}\otimes {\bf B_1}  -  {\bf A_2}\otimes {\bf B_2}.
\end{equation}

\noindent Tsirelson showed \cite{Tsirelson} that CHSH inequality (\ref{CHSH_LVM}) is
maximally violated by a multiplicative
factor $\sqrt{2}$ (Tsirelson's bound) on the basis of quantum mechanics. In fact, it is
true that $|Tr(\rho_{AB}{\cal B}_{CHSH})|\leq 2\sqrt{2}$ for all observables ${\bf A_1}$,
${\bf A_2}$, ${\bf B_1}$, ${\bf B_2}$, and all states $\rho_{AB}$. Increasing the
size of Hilbert spaces on either A and B sides would not give any advantage in the
violation of the CHSH inequalities. In general, it is not known how to calculate the best
such bound for an arbitrary Bell inequality, although several techniques have
been developed \cite{Toner}.

A good witness of useful correlations is, in many cases, the violation of a Bell inequality
by a quantum state. Although it is known that the
violation of an $N$-particle Bell-like inequality of some sort by an $N$-particle entangled state is not enough, {\it per se}, to prove genuine multipartite non-locality, it is the only approximation left in practice.

The first Bell inequality for $N=3$ qubits was provided by Mermin \cite{Mermin}. The Mermin inequality reads as $Tr(\rho {\cal B}_{Mermin}) \leq 2$, where ${\cal B}_{Mermin}$ is the Mermin operator

\begin{equation} \label{Mermin}
 {\cal B}_{Mermin}=B_{a_{1}a_{2}a_{3}} - B_{a_{1}b_{2}b_{3}} - B_{b_{1}a_{2}b_{3}} - B_{b_{1}b_{2}a_{3}},
\end{equation}

\noindent with $B_{uvw} \equiv {\bf u} \cdot {\bf \sigma} \otimes {\bf v} \cdot {\bf \sigma} \otimes {\bf w} \cdot {\bf \sigma}$
with ${\bf \sigma}=(\sigma_x,\sigma_y,\sigma_z)$ being the usual Pauli matrices, and ${\bf a_j}$ and ${\bf b_j}$ unit vectors
in $\mathbb{R}^3$. Notice that the Mermin inequality is maximally violated by Greenberger-Horne-Zeilinger (GHZ) states.

In the case of $N=4$ qubits, the first Bell inequality was derived by Mermin, Ardehali, Belinskii and Klyshko (MABK) \cite{MABK}. The MABK inequality reads as $Tr(\rho {\cal B}_{MABK}) \leq 4$, where ${\cal B}_{MABK}$ is the MABK operator

\begin{equation} \label{MABK}
\begin{split}
B_{1111}&-B_{1112} -B_{1121}-B_{1211}-B_{2111}-B_{1122}-B_{1212}\\
&-B_{2112}-B_{1221}-B_{2121}-B_{2211}+B_{2222}+B_{2221}\\
&+B_{2212}+B_{2122}+B_{1222},
\end{split}
\end{equation}

\noindent with $B_{uvwx} \equiv {\bf u} \cdot {\bf \sigma} \otimes {\bf v} \cdot {\bf \sigma} \otimes {\bf w} \cdot {\bf \sigma}\otimes {\bf x} \cdot {\bf \sigma}$
with ${\bf \sigma}=(\sigma_x,\sigma_y,\sigma_z)$ being the usual Pauli matrices. We shall define

\begin{equation} \label{MABKMax}
 B_{MABK}^{\max} \equiv \max_{\bf{a_j},\bf{b_j}}\,\,Tr (\rho {\cal B}_{MABK})
\end{equation}

\noindent as a measure for the nonlocality content for a given state $\rho$ of four qubits. ${\bf a_j}$ and ${\bf b_j}$ are unit vectors
in $\mathbb{R}^3$. MABK inequalities are such that they constitute extensions of the CHSH inequalities with the requirement that generalized
GHZ states maximally violate them.

The optimization \cite{jo1,jo2} is taken over the two observers' settings $\{{\bf a_j},{\bf b_j}\}$, which are real unit vectors in $\mathbb{R}^3$. We choose them to be
of the form $(\sin\theta_k \cos\phi_k,\sin\theta_k \sin\phi_k,\cos\theta_k)$. With this parameterization, the problem consists in finding
the supremum of $Tr(\rho{\cal B}_{MABK})$ over the ($N=4$) $\{k=1\dotsm 4N\}$ angles.

In the case of multiqubit systems, one must instead
use a generalization of the CHSH inequality to $N$ qubits. 
MABK inequalities are of such
nature  that they constitute extensions of older inequalities. 
To concoct an extension to the multipartite case, we shall
introduce a recursive relation \cite{GisinAntic} that will allow for more parties.
This is easily done by considering the operator

\begin{equation} \label{Biterative}
 B_{N+1}  \propto [(B_1+B_1^{\prime}) \otimes B_N + (B_1-B_1^{\prime}) \otimes B_N^{\prime}] ,
\end{equation}

 \noindent with $B_N$ being the Bell operator for N parties and $B_1={\bf v} \cdot {\bf \sigma}$,
 with ${\bf \sigma}=(\sigma_x,\sigma_y,\sigma_z)$ and ${\bf v}$ a real unit vector. The prime on the operator
 denotes the same expression but with all vectors exchanged. The concomitant maximum value

\begin{equation} \label{MABK_Nmax}
B_N^{\max} \equiv \max_{ \bf{a_j},\bf{b_j} }\,\,Tr (\rho {B_{N}})
\end{equation}

\noindent will serve as a measure for the non-locality content of a
given state $\rho$ of $N$ qubits if ${\bf a_j}$ and ${\bf b_j}$ are
unit vectors in $\mathbb{R}^3$. The non-locality measure
(\ref{MABK_Nmax}) is maximized by generalized GHZ states,
$2^{\frac{N+1}{2}}$ being the corresponding maximum value.

However, there exist other measures \cite{MABKnew} such as the Svetlichny inequalities \cite{Svetlichny} which serve the same purpose, having a similar structure extended to the $N$-partite scenario \cite{mauro1,mauro2}. They have been used in the literature as entanglement-like indicators \cite{Mauro}.

\subsection{The Turing machine}

A Turing machine \cite{Turing} has an infinite one-dimensional tape divided into cells. Traditionally we think of the tape as being horizontal with the cells arranged in a left-right orientation. The machine has a read-write head which is scanning a single cell on the tape. This read-write head can move left and right along the tape to scan successive cells. A table of transition rules will serve as the “program” for the machine.

In modern terms, the tape serves as the memory of the machine, while the read-write head is the memory bus through which data is accessed (and updated) by the machine. One very important aspect is that we shall rely on the Turing-computability of the cost function that maximizes the Bell inequality given a state $\rho_N$. As known, there exists an entire class of these problems which is termed “NP-complete” (non-deterministic polynomial time complete) because the computational effort
used to find an exact solution increases exponentially as the
total number of degrees of freedom of the problem rise.

\begin{figure}[htbp]
\begin{center}
\includegraphics[width=8.8cm]{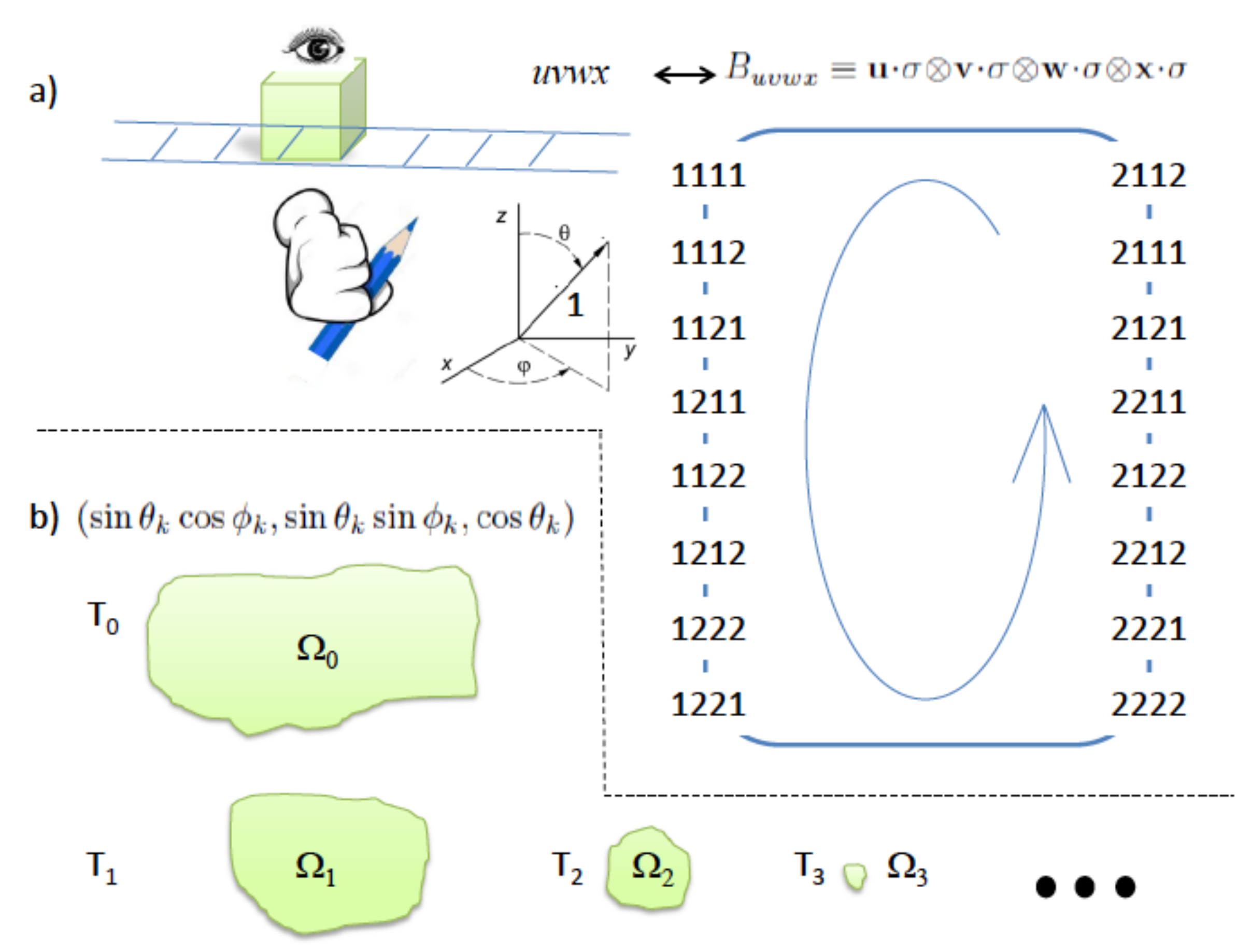}
\caption{(Color online) Fig. (a) Describes how a Turing machine computes the maximum violation of a Bell inequality for $N=4$ qubits within the framework of simulated annealing optimization. The number of terms to be evaluated by the machines grows exponentially as $4^{N-2}$. Fig. (b) Emphasizes the fact that as the temperature drops, the available set of $4N=16$ variables reduces until 
some precision on the cost function is reached. See text for details.}
\label{fig2}
\end{center}
\end{figure}

As a consequence, approximated or heuristic methods are required in practice for further
analysis. The most successful statistical method to date is the stochastic model of simulated annealing introduced by Kirkpatrick, Gelatt, and Vecchi \cite{kirkpatrick83}, that is, the
Metropolis Monte Carlo algorithm with a fixed temperature $T$ at each state of the 
annealing schedule. 
There exist other methods which are not of statistical nature, such as downhill/amoeba or gradient methods \cite{Avriel}, which involve finite differences when considering the corresponding function --a Bell inequality in our case-- in terms of all real variables involved.

\subsection{Results}

In either case --statistical or gradient-type method-- we can program the Turing machine in the same way, because after all it will undergo a Hamiltonian cycle changing the value of several parameters of the total function to be optimized and at every step of the procedure. Therefore, we choose a simulated annealing approach to the program. 

Regarding Bell inequalities, one can choose the MABK or the Svetlichny inequalities to maximize for a given state $\rho$. In either case, owing to (\ref{Biterative}), the number of individual terms grow exponentially as $4^{N-2}$, $N$ being the number of qubits. Let us then take the MABK inequalities. 

In order to illustrate how the Turing machine works, let us have the $N=4$ case. The total number of independent variables are $4N$, but what makes the computation hard is the number of constraints that we have. The situation is depicted in Fig. 2 (a). The Turing machine reaches the first term in the tensor expansion of the MABK inequality. It is free to move in space the unit vector of each party randomly, keeping in the memory that some vectors will have the same position in the next move for some parties. The temperature is high at $T_0$, which implies in Fig. 2 (b) that the domain of possible values for the variables $\Omega$ is broadly spread. Keep in mind, however, that every angle is reduced as follows: $\{\theta_i \rightarrow \theta_i \mod \pi,\psi_i \rightarrow \psi_i \mod 2\pi\}$. The machine then moves to the next term and performs similar operations accordingly. Finishing one cycle means visiting one after the other the entire $4^{4-2}=16$ sites. After that, the machine has to compute Tr($\hat{B_N} \hat{\rho_N}$), which is the cost function. Then, it starts the cycle anew with a different temperature $T_1$ (we can choose the temperature to decrease like 
$T(s)=T_0 e^{-\lambda s}$, with $s$ being the number of runs). As the temperature drops, the domain $\Omega_s$ shrinks as depicted in Fig. 2 (b). Thus, at every cycle the range of possible values for the $4N=8$ variables continuously decreases until we reach a desired precision, that is, the algorithm terminates when some stopping criterion is met. 

The basic algorithm is shown below.\\

\begin{footnotesize}
\begin{eqnarray}
1 &&\,\, \rho_N \,initial\, multiqubit\, (N) \,state\, given\cr
2 &&\,\, T \leftarrow T_0\cr
3 &&\,\, {\bf repeat \,until} \,stopping\, criterion \,is \,met\cr
4 &&\,\, ~~~~~~~{\bf repeat\,} 4^{N-2} \,times \cr
5 &&\,\, ~~~~~~~~~~~~~~orientate \,N\, unit \,vectors \cr
6 &&\,\, ~~~~~~~~~~~~~~move \,to \,the \,next \,term \,within \,the \,set \cr
7 &&\,\, ~~~~~~~{\bf endrepeat}  \cr
8 &&\,\, ~~~~~~~super\, operator \,B_N \,\leftarrow \,add \,all \,terms  \cr
9 &&\,\, ~~~~~~~calculate \,Tr(B_N \rho_N) \cr
10&&\,\, ~~~~~~~T \,\leftarrow \,T_0 e^{-\lambda s} \cr
11&&\,\, {\bf endrepeat} \cr
12&&\,\, return \,B_N^{\max} \equiv Tr(B_N \rho_N)\cr
\end{eqnarray}
\end{footnotesize}

If we do not want to specify a method in solving the optimization, we can rewrite the algorithm as:

\begin{footnotesize}
\begin{eqnarray}
1 &&\,\, \rho_N \,initial\, multiqubit\, (N) \,state\, given\cr
2 &&\,\, {\bf repeat \,until} \,stopping\, criterion \,is \,met\cr
3 &&\,\, ~~~~~~~{\bf repeat\,} 4^{N-2} \,times \cr
4 &&\,\, ~~~~~~~~~~~~~~perform \, operations\, on\, N\, parties'\, observables  \cr
5 &&\,\, ~~~~~~~~~~~~~~register \, in\, memory \cr
6 &&\,\, ~~~~~~~~~~~~~~move \,to \,the \,next \,term \,within \,the \,set \cr
7 &&\,\, ~~~~~~~{\bf endrepeat}  \cr
8 &&\,\, ~~~~~~~super\, operator \,B_N \,\leftarrow \,add \,all \,terms  \cr
9 &&\,\, ~~~~~~~calculate \,Tr(B_N \rho_N) \cr
10&&\,\, {\bf endrepeat} \cr
11&&\,\, return \,B_N^{\max} \equiv Tr(B_N \rho_N)\cr
\end{eqnarray}
\end{footnotesize}\newline

\noindent Every cycle contains at least $4^{N-2}$ visits, and the best computation of line 9 in the previous algorithm for the Turing machine is of $O(d^3)$, 
where $d$ is the dimension of the square matrices ($d=2^N$ in our case) being multiplied \cite{matrixproductMilan}. Therefore, we have an undefined number of times (at least two) $\times 4^{N-2} \times (2^N)^3$ number of steps required to obtain 
$B_N^{\max}$. In other words, {\it at least} we require $O(2^{5N})$ steps to solve the problem which, 
in view of the aforementioned result, clearly becomes NP with increasing number of parties. This is precisely the desired outcome: the computation of the maximum value of a Bell inequality requires a computational effort which grows exponentially with the number of parties $N$ involved.

\section{Conclusions}

Based on the iterative structure of the extension of Bell inequalities to the multiqubit case, we have shown that the maximization of the usual Bell inequalities employed in the literature (except the ones for probabilities, as in \cite{Gisin3}), an operation performed by a Turing machine, constitutes a NP-problem. 
This results somehow express the fact that, regarding nonlocality as a good resource for quantifying quantum correlations other than entanglement, the concomitant optimization becomes non-tractable for high number of qubits. Furthermore, even the fact of cheeking the plain violation of an inequality for a state $\rho_N$, which implies Tr($B_N \rho_N$), is of $O(2^{3N})$, that is, it is limited in practice to a few number of qubits.    

\section*{Acknowledgements}

J. Batle acknowledges fruitful discussions with J. Rossell\'o, Maria del Mar Batle and Regina Batle. 
R. O. acknowledges support from High Impact Research MoE Grant UM.C/625/1/HIR/MoE/CHAN/04 from the
Ministry of Education Malaysia.

\end{document}